\documentclass[useAMS,usenatbib,usegraphicx]{mn2e}

\usepackage{times}

\newcommand\hi{\mbox{H\,{\sc i}}}

\title[The Spatially-Resolved Star Formation History of the M31 Outer Disc]{The
Spatially-Resolved Star Formation History of the M31 Outer Disc\thanks{Based on
observations made with
    the NASA/ESA Hubble Space Telescope, obtained at the Space Telescope
    Science Institute, which is operated by the Association of
    Universities for Research in Astronomy, Inc., under NASA contract
    NAS5-26555. These observations are associated with programmes
    GO-9458 and GO-13018.}}
\author[E.~J.\ Bernard et al.]{%
Edouard J. Bernard,$^1$\thanks{E-mail: ejb@roe.ac.uk},
Annette M. N. Ferguson,$^1$
Scott C. Chapman,$^2$
Rodrigo A. Ibata,$^3$
\newauthor Mike J. Irwin,$^4$
Geraint F. Lewis,$^5$
Alan W. McConnachie$^6$\\
 $^{1}$SUPA, Institute for Astronomy, University of Edinburgh, Royal
   Observatory, Blackford Hill, Edinburgh EH9 3HJ, UK\\
 $^{2}$ Department of Physics and Atmospheric Science, Dalhousie University,
   Coburg Road, Halifax B3H1A6, Canada\\
 $^{3}$ Observatoire de Strasbourg, 11, Rue de l'Universit\'e,
   F-67000 Strasbourg, France\\
 $^{4}$ Institute of Astronomy, Madingley Road, Cambridge CB3 0HA, UK\\
 $^{5}$ Sydney Institute for Astronomy, School of Physics, A28, University
   of Sydney, NSW 2006, Australia\\
 $^{6}$ NRC Herzberg Institute of Astrophysics, 5071 West Saanich Road,
   Victoria, BC, V9E 2E7, Canada}

\begin{document}

\date{Accepted --. Received --; in original form --}

\pagerange{\pageref{firstpage}--\pageref{lastpage}} \pubyear{2015}

\maketitle

\label{firstpage}

\begin{abstract}
 We present deep {\it Hubble Space Telescope} Advanced Camera for
 Surveys observations of the stellar populations in two fields lying
 at 20 and 23~kpc from the centre of M31 along the south-west
 semi-major axis. These data enable the construction of
 colour-magnitude diagrams reaching the oldest main-sequence
 turn-offs ($\sim$13~Gyr) which, when combined with another field at
 25~kpc from our previous work, we use to derive the first precision
 constraints on the spatially-resolved star formation history of the
 M31 disc. The star formation rates exhibit temporal as well as
 field-to-field variations, but are generally always within a factor
 of two of their time average. There is no evidence of inside-out
 growth over the radial range probed. We find a median age of
 $\sim$7.5~Gyr, indicating that roughly half of the stellar mass in
 the M31 outer disc was formed before z~$\sim 1$. We also find that
 the age--metallicity relations (AMRs) are smoothly increasing from
 [Fe/H]$\simeq$-0.4 to solar metallicity between 10 and 3~Gyr ago,
 contrary to the flat AMR of the Milky Way disc at a similar number
 of scale lengths. Our findings provide insight on the roles of
 stellar feedback and radial migration in the formation and evolution
 of large disc galaxies.
\end{abstract}


\defcitealias{ber12}{Paper~I}
\defcitealias{ber15}{Paper~II}


\section{Introduction}

Disc galaxies account for a sizeable fraction of the stellar mass in
the Universe, yet the details of their formation and evolution are
still greatly debated. While much progress has been made in simulating
realistic disc galaxies within a cosmological framework
\citep[e.g.][]{aum13,mar14}, many issues remain unresolved. Strong
stellar feedback at early epochs is often included in hydrodynamical
simulations in order to suppress high redshift star formation,
postponing the formation of extended discs until relatively recent
times (z$\la$1 or $\sim$8 Gyr) when the bulk of the merging is over
\citep[e.g.][]{sti13}. This scenario naturally results in the
inside-out growth of discs, predicting radial age gradients and young
mean stellar ages at large radii. Direct observations of old and
intermediate age stars in outer discs are required to shed light on
the efficacy and role of early feedback \citep[e.g.][]{fer01, wil13}.

Once in place, various processes may rearrange the spatial
distribution of the stellar populations within galactic discs. For
example, it has been shown that stars can undergo large radial
excursions within discs due to scattering off transient spirals and
other features \citep[e.g.][]{sel02,min11}. By efficiently mixing
stars of different ages and metallicities over kiloparsec scales, this
process could lead to a smearing of the star formation history (SFH)
and age--metallicity relation (AMR) at a given radius, therefore
complicating the interpretation of the resolved fossil record.
Determining the importance of this process in real galaxy discs is
obviously of paramount importance.

To explore these issues, we have obtained deep {\it Hubble Space
Telescope (HST)} pointings with the Advanced Camera for Surveys
(ACS) in order to derive the detailed SFH and AMR at several locations
in the outskirts of M31. The SFHs that we calculated for the first 14
fields revealed a global burst of star formation in the M31 outer disc
roughly 2~Gyr ago, and allowed us to determine the nature and origin
of the various substructures in the outer disc and inner halo
\citep[hereinafter Papers~I and II, respectively]{ber12,ber15}. In
this Letter, we present the SFHs calculated for three very deep fields
located between 20 and 25~kpc from the centre of M31 along the
south-west semi-major axis. Based on colour-magnitude
diagrams (CMDs) that reach back to the oldest main-sequence turn-offs
(MSTO), these data provide the first precision view of the
spatially-resolved SFH in {\it any} disc galaxy and lead to a number
of interesting conclusions about the assembly history of the M31 outer
disc. The paper is structured as follows: in Section~\ref{obs}, we
present the observations and the data reduction steps, while the
CMD-fitting method is briefly described in Section~\ref{sfh}. The
resulting SFHs and our interpretation of the results are discussed in
Section~\ref{interp}.

\begin{figure}
\includegraphics[width=8cm]{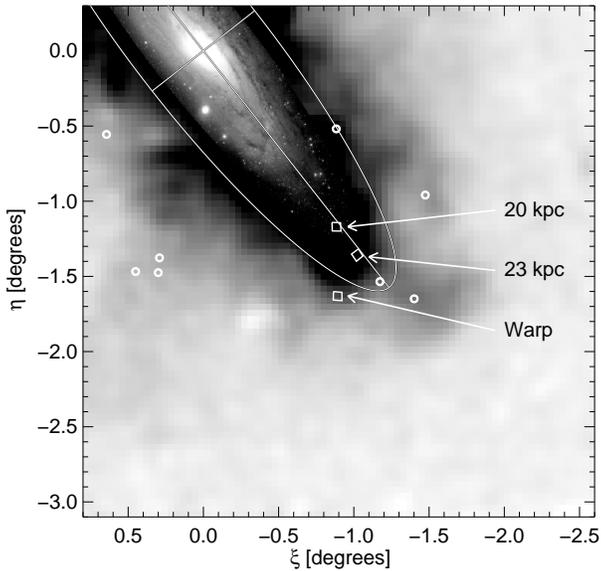}
\caption{Location of our {\it HST/ACS} pointings superimposed onto the
  INT/WFC map of M31's inner halo \citep{irw05}, showing the
  distribution of evolved giant stars around M31. The three fields
  analyzed in this work are shown as open squares, while the open
  circles represent some of the fields studied in Papers I and II. The
  ellipse has a semimajor axis of 2$\degr$ (27~kpc) and represents an
  inclined disc with $i$ = 77\fdg5 and position angle of
  38\fdg1.}\label{fig:map}
\end{figure}


\section{Observations and Data Reduction}\label{obs}

The 20 and 23 kpc fields were observed with the ACS onboard the {\it
HST} (proposal GO-13018, P.I.: A.\ Ferguson) between 2013 May and
August. As the goal of the observations was to study the history of
the outer thin disc, the fields were placed very close to the
semi-major axis. When combined with the {\it Warp} field\footnote{This
field lies in the southern stellar warp, which bends strongly away
from the major axis. As a result, its true radius in the disc could
be somewhat larger than its projected radius.} analysed in
\citetalias{ber12}, the three fields probe a range of galactocentric
distance from 20.1 to 25.4 kpc along the southwest thin disc,
corresponding to $\sim$3.8 to 4.8 scalelengths in the M31 disc
assuming $R_{\rm{d}} = 5.3 \pm 0.5\ \rm{kpc}$ \citep{cou11}. In terms
of radial scalelengths, we note that the radius of our innermost field
is comparable to that of the solar neighbourhood\footnote{The Sun is
located $\sim$3.6 scalelengths from the Galactic centre, assuming
$\rm{R}_{d,MW}\sim2.2$~kpc \citep[e.g.][]{bov13} and
$\rm{R}_0=8$~kpc \citep[e.g.][]{mal13}.}. Our fields are shown as
open squares in Fig.~\ref{fig:map}, overplotted on a map of the
surface density of red giant branch (RGB) stars from the INT/WFC
survey of \citet{irw05}. The {\it HST} field lying on the major axis
just inside the 27~kpc ellipse is the {\it Outer Disc} field studied
in \citetalias{ber12}, for which we could not calculate the SFH
because of strong differential reddening from dust within M31; the
other open circles represent some of the inner halo and substructure
fields for which SFHs were derived in \citetalias{ber15}. While the
{\it Warp} was observed for 10 {\it HST} orbits, we obtained 13 orbits
per field for each of the new fields in order to compensate for the
effect of higher stellar density and background flux. A detailed
summary of observations including coordinates, foreground color
excess, radial distances, and exposure times is given in
Table~\ref{tab1}.

The photometry and artificial star tests were carried out following
identical methods as those described in \citetalias{ber12}. The only
difference here is that we took advantage of the images corrected for
the charge transmission inefficiency (`{\tt FLC}') provided by the
{\it HST} pipeline \citep[e.g.][]{and10,ube12}. The stellar photometry
was carried out on the individual exposures with the standard {\sc
daophot/allstar/allframe} suite of programs \citep{ste94}. They were
then cleaned of non-stellar objects by applying cuts on the
photometric parameters given by {\sc allframe}, namely the photometric
uncertainty ($\sigma \leq 0.3$) and the sharpness, describing how much
broader the profile of the object appears compared to the profile of
the PSF ($|{\tt SHARP}|\leq 0.3$). After cleaning, the CMDs contain
353\,918 and 296\,731 stars, respectively, compared to 185\,335 in the
{\it Warp} with the same quality cuts. The final photometry was
calibrated to the VEGAMAG system following the prescriptions of
\citet{sir05} and the revised ACS
zeropoints\footnote{http://www.stsci.edu/hst/acs/analysis/zeropoints.},
then to absolute magnitude by taking into account the distance to M31
\citep[(m$-$M)$_0$=24.47; e.g.][]{mcc05}, and the foreground reddening from
the \citet{sch98} extinction maps together with the updated coefficients
from \citet{sch11}. Note that the CMD-fitting technique used in this paper
minimizes the impact of the uncertainties in distance, mean reddening,
and photometric zero-points on the solutions by shifting the observed
CMD with respect to the artificial CMD; therefore, any small errors in
these quantities do not affect the results.

\begin{table}
 \centering
 \begin{minipage}{85mm}
  \caption{Summary of Observations\label{tab1}}
  \begin{tabular}{@{}lcc@{}}
  \hline
  Field                                                          &  20 kpc                              & 23 kpc               \\
  \hline
  RA (J2000)                                                     &  00:38:05.8                          &  00:37:23.8         \\
  Decl. (J2000)                                                  & +40:05:35                            & +39:54:00            \\
  $R$\footnote{Galactocentric distance.} (kpc)                   & 20.1                                 & 23.3                 \\
  E(B$-$V)\footnote{Values from \citet{sch11}.}                  & 0.053                                & 0.055                \\
  Dates                                                          & 2013 May 28--Jun 2                   & 2013 Jul 31--Aug 10  \\
  t$_{F606W}$\footnote{Individual and total exposure times.} (s) & 4$\times$1149+8$\times$1245=14\,556  & 4$\times$1145+8$\times$1240=14\,500  \\
  t$_{F814W}$$^c$ (s)                                              & 4$\times$1149+10$\times$1245=17\,046 & 4$\times$1145+10$\times$1242=17\,000 \\
  \hline\vspace{-10mm}
\end{tabular}
\end{minipage}
\end{table}


\begin{figure*}
\includegraphics[width=14.8cm]{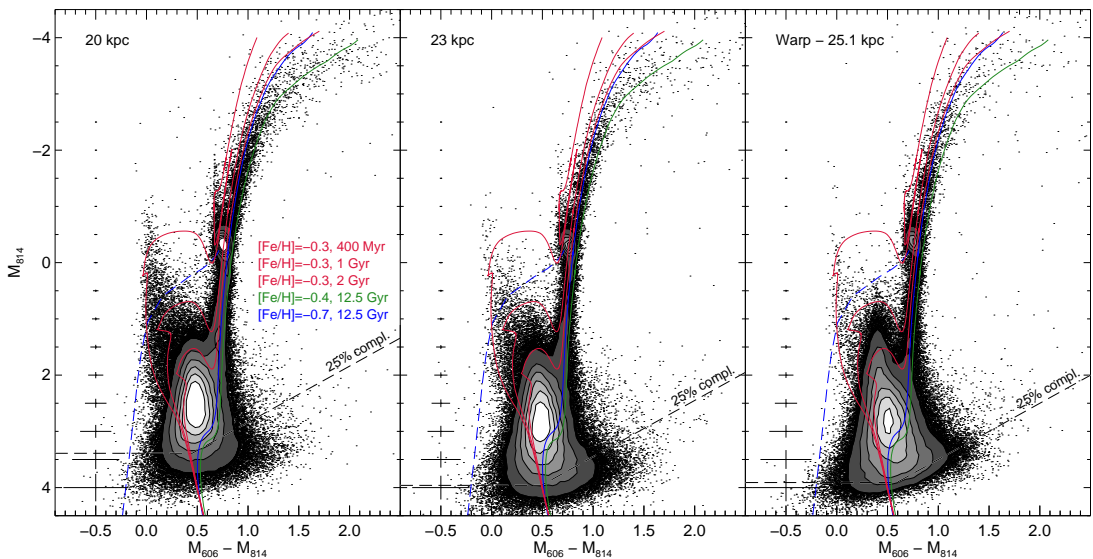}
\caption{CMDs for the {\it 20 kpc} (left), {\it 23 kpc} (middle), and
  {\it Warp} (right) fields, where selected isochrones (solid lines)
  and a ZAHB (long-dashed line) from the BaSTI library \citep{pie04}
  are overlaid and labelled in the inset. Only 120\,000 stars
  per CMD are shown. The error bars show the mean photometric errors
  as a function of magnitude. The contour levels correspond to [20,
  40, 60, 80, 100, 120]$\times 10^3 {\rm \ stars \ mag}^{-2}$. The
  25\% completeness limit from the artificial stars tests is shown by
  the dashed lines.}\label{fig:cmds}
\end{figure*}

While great care was taken to place the fields in areas of low dust
extinction by using the \hi\ maps from \citet{bra09}, we found that the
{\it 23~kpc} field is affected by mild differential reddening due to
dust in M31. We used the same method as in \citetalias{ber12} for the
{\it Outer Disc} field to quantify it. We were not able to correct the
areas with measureable differential reddening, but found that $\sim$40
percent of the ACS field was not affected. In the following we only
use the $\sim$120\,000 stars that reside in the unaffected areas.

The resulting CMDs for the two new fields plus
the {\it Warp} are shown in Fig.~\ref{fig:cmds}; only 120\,000 stars
per CMD are shown to ease the comparison. Selected isochrones and a
theoretical zero-age horizontal-branch (ZAHB) from the BaSTI library
\citep{pie04} are shown as labeled in the inset, both to indicate the
location of the MSTOs and as a guide for comparing the CMDs. The 25\%
completeness limit from the artificial star tests is represented as
dashed lines. It shows that while the {\it 23~kpc} and {\it Warp}
field have the same photometric depth, the {\it 20~kpc} field is
slightly shallower due to the higher stellar density and sky
background at the time of the observations. In terms of stellar
populations, the main differences between these fields are the
gradually less populated main-sequence (at $M_{606}-M_{814}\sim0$
and M$_{814}\la2$) with increasing radius, and the RGB of the
{\it Warp} above M$_{814}\sim-1$ being on average bluer by 0.05 mag
than in the other fields.
The {\it Warp} also contains a densely populated plume
of stars at $M_{606}-M_{814}\sim0.4$ and M$_{814}\sim1.5$ that
corresponds to a $\sim$2~Gyr old stellar population (see below), and
which is not as prominent in the other fields.


\section{Star formation history calculation}\label{sfh}

The SFHs were calculated using the technique of synthetic CMD fitting,
which provides a detailed and robust quantitative estimate of the star
formation rate (SFR) and metallicity ([Fe/H]) at a given time. This
involved fitting the observed data with synthetic CMDs to extract the
linear combination of simple stellar populations (SSP) -- i.e.\ each
with small ranges of age ($\leq$1~Gyr) and metallicity ($<$0.25~dex)
-- which provides the best fit; the amplitudes of the required SSPs
give the rates of star formation as a function of age and metallicity.

\begin{figure*}
\includegraphics[width=5.8cm]{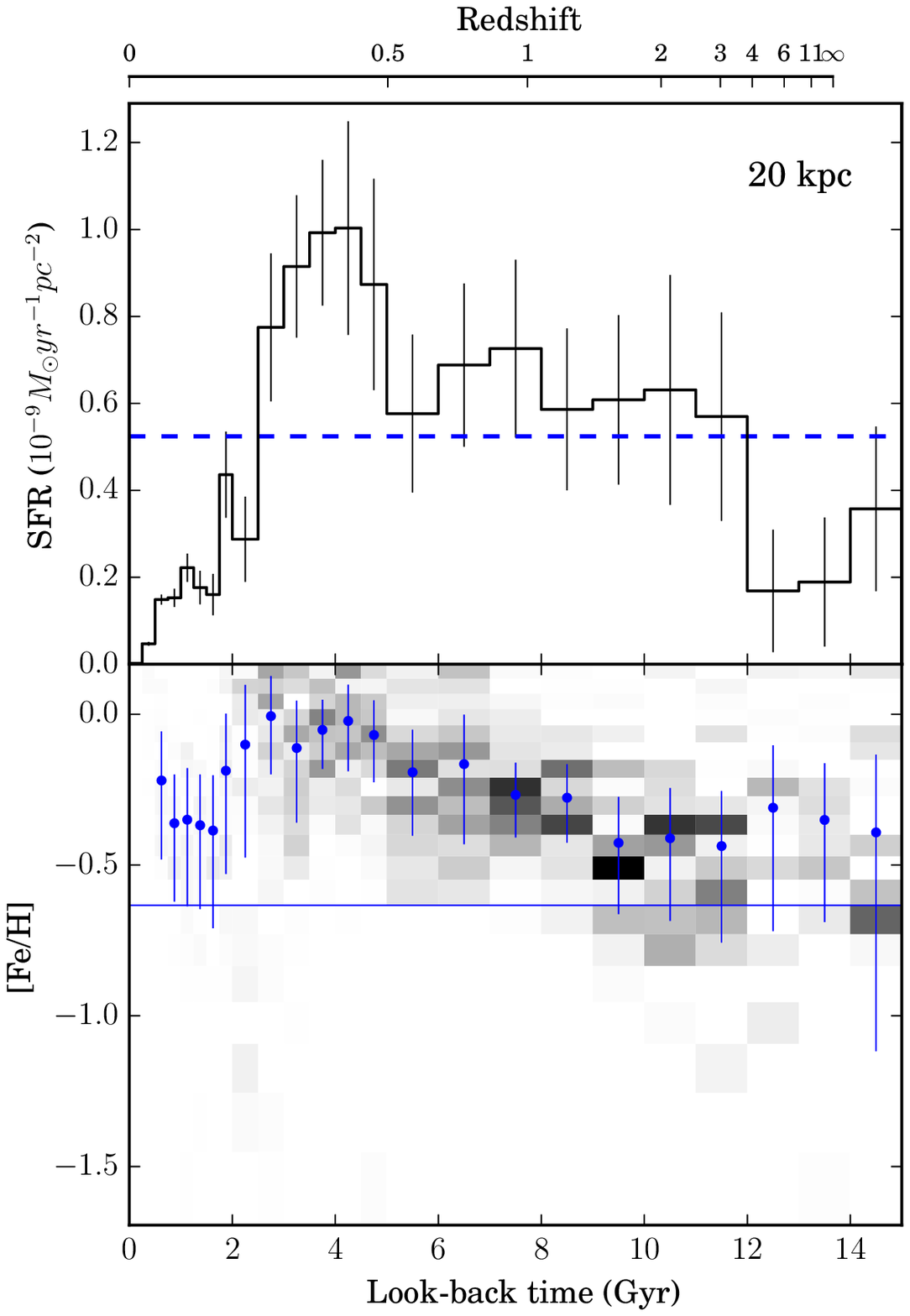}
\includegraphics[width=5.8cm]{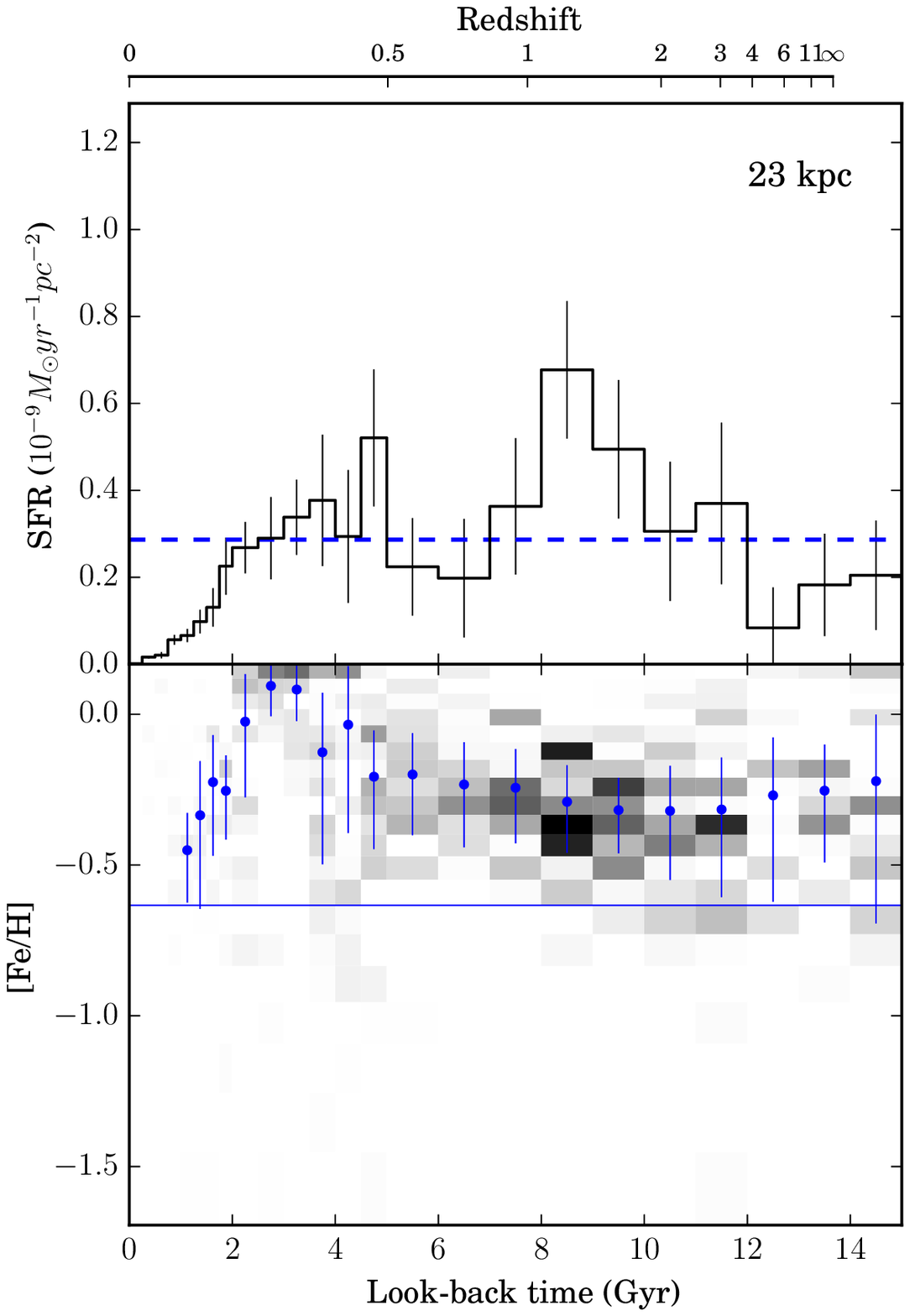}
\includegraphics[width=5.8cm]{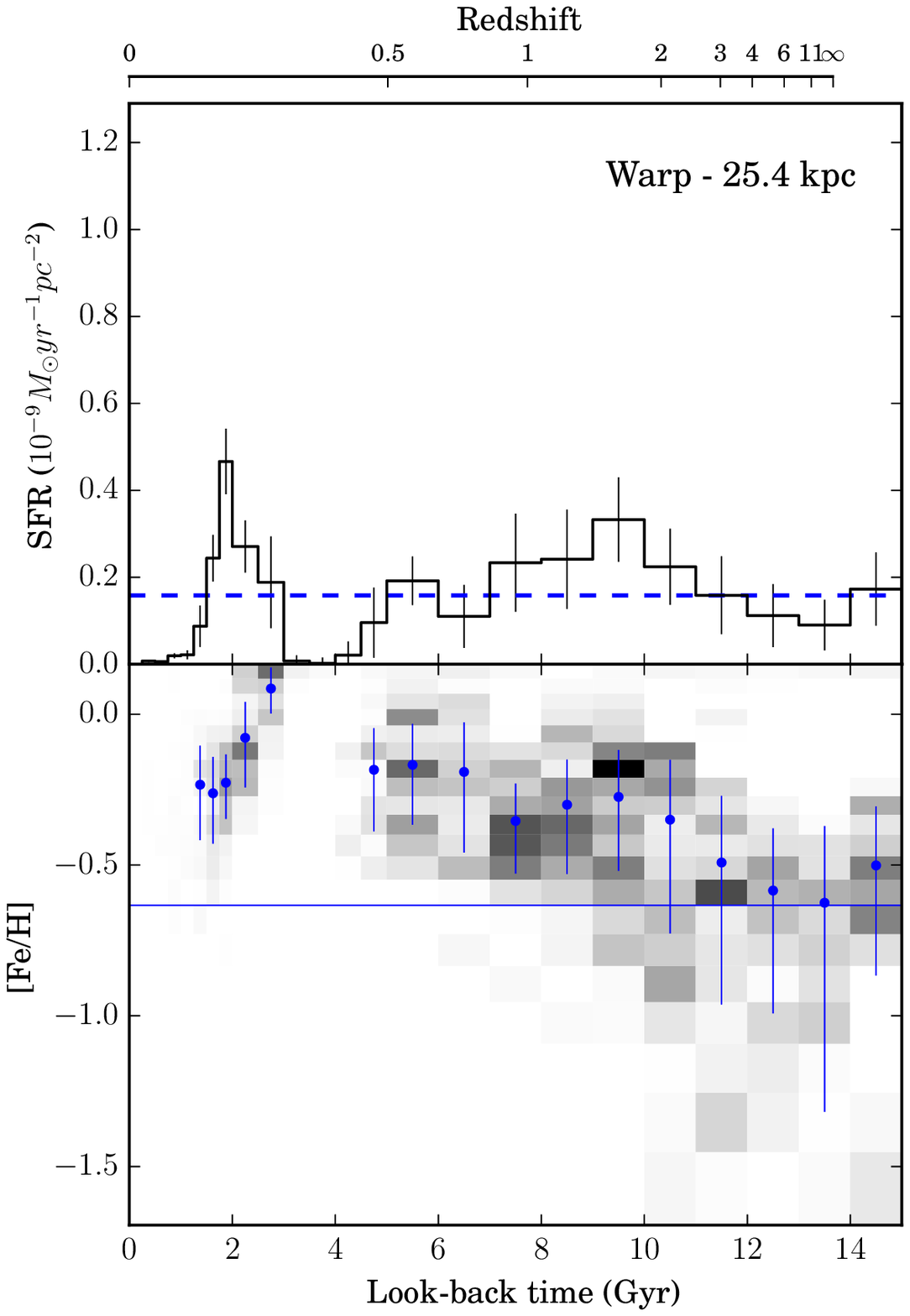}
\caption{Evolution of the SFR (top) and metallicity (bottom) as a
  function of time for the three fields, displayed on the same
  absolute scale. The dashed lines in the top panels show the average
  SFR over the age of the disc. In the bottom panels, the grayscale is
  proportional to the stellar mass formed in each bin. The filled
  circles and error bars show the mean metallicity and standard
  deviation in the age bins representing at least 1~percent of the
  total mass of stars formed; the solid blue line serves to guide the
  eye.}\label{fig:3sfhs}
\end{figure*}

The methodology is identical to that used in Papers~I and II, although
we are now using a new algorithm developed in Python by one of us
(EJB). The only notable difference is that the goodness of fit is
measured through a Poisson equivalent of $\chi^2$ \citep[see][]{dol02}
rather than the modified $\chi^2$ statistics of \citet{mig99}.
Extensive testing has shown that it produces results that are
virtually indistinguishible from those obtained with the algorithm
used in our previous papers \citep[{\sc iac-pop/minniac};
e.g.][]{hid11}. This can be verified by comparing the SFH of the {\it
Warp} that was calculated using the new algorithm (see
Fig.~\ref{fig:3sfhs}) with Fig. 6 in \citetalias{ber12}. The
advantages of the new approach are twofold: the algorithm is at least
an order of magnitude faster -- a necessary efficiency in view of the
ever-increasing size of star catalogues -- and it is significantly
easier to use. We therefore refer the interested reader to Papers~I
and II for a detailed description of the assumptions and fitting
procedure.

\section{Results and discussion}\label{interp}

The resulting SFHs are presented in Fig.~\ref{fig:3sfhs}, where the
top and bottom panels show the evolution of the SFR and of the
metallicity, respectively. The redshift scale was constructed assuming
the WMAP7 cosmological parameters \citep{jar11}. In all fields, we
find that star formation began early on (z $\ga2$) and occurred more
or less continuously across the history of the disc. This indicates
that M31 was already a large disc galaxy at high redshift and could be
a direct descendent of the large disc systems identified in distant
redshift studies \citep[e.g.][]{for09,van13}. The SFHs exhibit complex
temporal variations; aside from the strong decline over the last
2~Gyr, these variations are not clearly correlated from field to
field. Nonetheless, the SFR in a given field remains within a factor
of $\sim2$ of its time average, indicating that very strong bursts of
star formation have not dominated the mass build-up in the stellar
disc. As expected from the decreasing stellar density with increasing
galactocentric radius, the average SFR decreases by a factor $\sim$3.3
over the radial range spanned by our data (roughly one
scalelength).

A particular feature that stands out in the SFH of the {\it Warp} is
the burst of star formation $\sim$2~Gyr ago, associated with the onset
of a decrease in global metallicity, that we first detected in
\citetalias{ber12} and later found to be widespread in more distant
(R$\ga25$ kpc) fields in M31 \citepalias{ber15}. Interestingly, while
the decrease in [Fe/H] is clearly measured in the two new fields
presented here, we do not detect any clear enhancement in the
SFR at that epoch. The 20 kpc field does show an enhanced
period of star formation activity but this is not as pronounced as the
{\it Warp} burst, and occurs at earlier epochs (3-5 Gyr), so it is not
clearly related. We thus conclude that either the 2 Gyr burst was
confined to the very outer disc, or that it had much greater
significance there. These findings appear to be at odds with SFHs
calculated by \citet{wil15} in the northern inner disc of M31, which
show a ubiquitous strong burst 2-4 Gyr ago in fields ranging from
3--20 kpc. The reason for this discrepancy is presently not clear,
although we note that their result is largely based on modelling
evolved stars (red clump and RGB) as age indicators, which is not as
robust as modelling the main sequence turn-offs in our much 
deeper data.

\begin{figure}
\includegraphics[width=8cm]{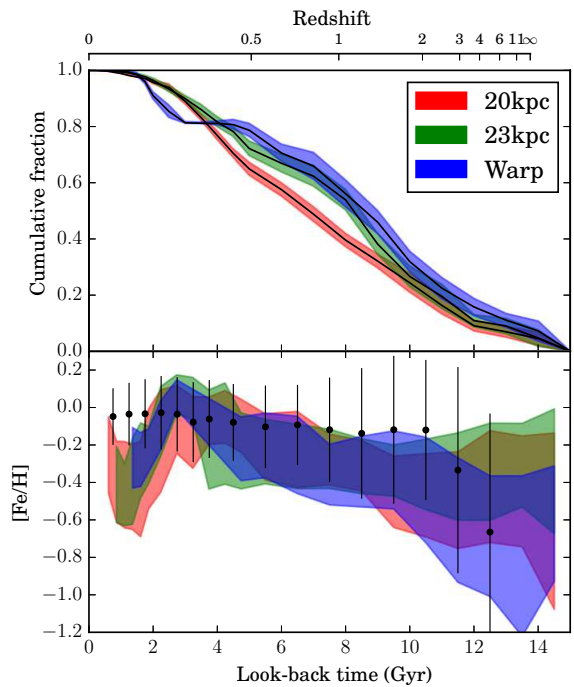}
\caption{Comparison of the SFHs of the three fields, showing the
  evolution of the cumulative mass fraction (top) and metallicity
  (bottom), as labeled in the inset. The integrated stellar mass
  density in each field at the present-day is $3.9, 2.1\ \rm{and}\ 1.2
  \times 10^6$M$_{\odot}$ kpc$^{-2}$ for the 20 kpc, 23 kpc, and {\it
  Warp} fields, respectively. In the bottom panel, the contours
  correspond to the 1-$\sigma$ envelope for the age bins representing
  at least 1~percent of the total mass of stars formed, shown as blue
  error bars in Fig.~\ref{fig:3sfhs}. The black filled circles with
  error bars show the AMR of the solar neighbourhood (mean metallicity
  and standard deviation) from a reanalysis of the Geneva-Copenhagen
  Survey \citep{cas11} for comparison.}\label{fig:sfhs}
\end{figure}

To facilitate comparison, Fig.~\ref{fig:sfhs} combines the SFHs of our
three fields, where the top and bottom panels show respectively the
cumulative mass fraction and the AMR as a function of time. The top
panel reaffirms that even though the details of the stellar mass
growth vary between the fields, all three show significant star
formation at $z \ga 1$. Combining all three fields, the median
stellar age in the outer disc fields is $\sim 7.5$ Gyr. To test the
reliability of these old ages, we ran constrained SFH calculations
where only model populations younger than a given age (8~Gyr, 10~Gyr)
were allowed. While this affected the ability of the algorithm to fit
the observed CMD, the median age of the populations did not change.
With the caveat that our fields span only one scalelength in radius,
there is no signal of inside-out growth in our data. The epoch at
which 50\% of the present-day stellar mass was in place is 6.9, 8.2,
and 8.6 Gyr for the 20 kpc, 23 kpc, and {\it Warp} fields,
respectively. The mild trend of older mean age for increasing radius
in M31 is in contrast to some recent integrated light analyses of
local galaxies \citep[e.g.][]{per13}.

The bottom panel of Fig.~\ref{fig:sfhs} shows that, for the past
9-10~Gyr, the AMRs of the three fields are very similar within the
uncertainties. All start from a moderate level of pre-enrichment
and all exhibit a clear decline in the last 3 Gyr. However, we find
that the older populations of the {\it Warp} are considerably more
metal-poor than in the 20 and 23~kpc fields, consistent with the
slightly bluer RGB in that field (see Fig.~\ref{fig:cmds}). This
variation is in keeping with the mild abundance gradient that has been
measured in this part of the M31 disc \citep[e.g.][]{kwi12}.

However, the most striking feature of these AMRs is the relatively
tight, smoothly increasing metallicity between $\sim$10 and 3~Gyr ago.
This is in stark contrast with the flat AMR of the solar neighbourhood
over the same age range \citep[e.g.][]{hay08,cas11}. This comparison
is relevant since our 20~kpc field in M31 is located at roughly the
same number of radial scalelengths as the Sun in the Milky Way. We
compare the Geneva-Copenhagen Survey \citep[GCS;][]{cas11}
AMR\footnote{Where the age is their ``BaSTI Max likelihood age''.}
with ours in the bottom panel of Fig.~\ref{fig:sfhs}; it shows that
their analysis revealed no metallicity increase in the past 10~Gyr,
compared to the enhancement of about 0.4~dex that we have measured in
the disc of M31. We note that more recent works on the AMR of the
solar neighbourhood based on high resolution spectra
\citep{hay13,berg14} do find a clear metallicity increase before about
8~Gyr ago, but a similarly flat AMR since $z\sim1$.

The difference between the AMRs of the M31 and the Milky Way discs may
indicate different formation and/or subsequent evolution histories.
The flattening and high dispersion of the AMR in the solar
neighbourhood is often ascribed to effects of radial migration in the
disc due to non-axisymmetric perturbations of the gravitational
potential \citep[e.g.][]{sel02}, this region containing stars born in
a wide range of galactocentric distances. The increasing AMR in the
outer disc of M31, as well as the complex and differing SFH behaviours
observed in the three fields, suggests that migration in this galaxy,
or at least this part of the galaxy, has been less efficient than in
the solar neighbourhood. For example, it is likely that the old, most
metal-poor component of the {\it Warp} would have had sufficient time
to contaminate the {\it 23~kpc} field had migration been significant.
The different migration efficiencies may be related to the different
dynamical properties of the M31 and Milky Way discs: \citet{ver14}
have shown that the efficiency of radial migration is a strong
function of the vertical velocity dispersion, the stars spending more
time near the disc plane being more affected by the non-axisymmetric
perturbations. The measurements of \citet{dor15} in the disc of M31
indicate a velocity dispersion at a given age that is more than twice
as large as in the Milky Way (see also \citet{col11}), potentially
providing an explanation for the lack of migration. The
old age of the M31 outer disc provides further indirect support for
this idea, since disc galaxies observed at z$\sim2$ are often highly
turbulent \citep[e.g.][]{for09}.

In summary, our analysis of deep CMDs that reach back to the oldest
MSTOs have enabled the first precision constraints on the radial
variation of the star formation history in the M31 outer disc. Roughly
half of the stellar mass in our fields was formed before z~$\sim 1$,
indicating that the extended M31 disc was in place and actively
forming stars at high redshift. Stellar feedback was not strong enough
to significantly delay the formation of the outer disc. Our results
also suggest that radial migration has not played a major role in
rearranging the stellar populations at large radii in M31. Notably, a
well-defined AMR is seen in all fields, in marked contrast the flat
AMR seen in the Milky Way at similar galactocentric distances. Our
findings provide new insight on processes relevant for modelling the
evolution of disc galaxies, issues which we will explore in more
detail in future work.


\section*{Acknowledgments}

This work was supported by a consolidated grant from STFC.


\label{lastpage}

\end{document}